\documentclass[sigplan,nonacm]{acmart}

\usepackage{amsmath}
\usepackage{graphicx}
\usepackage{xcolor}
\usepackage{booktabs}
\usepackage{listings}
\microtypesetup{protrusion=true,expansion=false}
\usepackage[htt]{hyphenat} 
\setcopyright{none}

\newcommand{\addfigcols}[3]{
    \begin{figure*}[hbt!]
        \centering
        \includegraphics[width=0.8\linewidth,height=\textheight,keepaspectratio]{#3}
        \vspace{-0.15in}
        \caption{\footnotesize{#2}}
        \label{#1}
        \vspace{-0.10in}
    \end{figure*}
}

\def\eqref#1{Equation~\ref{#1}}

\definecolor{mlirkw}{rgb}{0.00,0.35,0.62}
\definecolor{mlircom}{rgb}{0.40,0.40,0.40}
\definecolor{mlirtype}{rgb}{0.00,0.55,0.25}
\lstdefinelanguage{MLIR}{
  morekeywords={func,return,module},
  morekeywords=[2]{ftqc,arith,scf,i1},
  sensitive=true,
  morecomment=[l]{//},
  morestring=[b]",
}
\newcommand{\steane}{$[\![7,1,3]\!]$}

\newif\ifdraft
\drafttrue

\ifdraft
\newcommand{\todo}[1]{\textcolor{green}{\textbf{TODO:} #1}}
\newcommand{\xcomment}[1]{\textcolor{green}{\textbf{TODO:} #1}}

\newcommand{\elaine}[1]{\textcolor{magenta}{#1}}
\newcommand{\issue}[1]{\textcolor{red}{\textbf{Issue:}#1}}
\else
\newcommand{\todo}[1]{}
\newcommand{\xcomment}[1]{}

\newcommand{\elaine}[1]{}
\newcommand{\issue}[1]{}
\fi

\def\BibTeX{{\rm B\kern-.05em{\sc i\kern-.025em b}\kern-.08em
    T\kern-.1667em\lower.7ex\hbox{E}\kern-.125emX}}

\begin{document}

\title{A Compiler for Fault Tolerant Quantum Computing}

\begin{abstract}
Fault-tolerant quantum computing (FTQC) needs compiler intermediate representations (IRs) that preserve error-correcting-code choices, syndrome information, and non-Clifford resource requirements through analysis, simulation, and hardware lowering. We report practical experience building \texttt{FTQC}, an open-source Multi-Level Intermediate Representation (MLIR) dialect that makes logical qubits, syndromes, magic states, and Pauli frames typed static single assignment (SSA) values. Its error-correction-aware type system, two-level lowering design, and IR-resident resource analysis provide one inspectable pipeline to Open Quantum Assembly Language~3 (OpenQASM~3), the Quantum Intermediate Representation (QIR), Stim, and IQM hardware. Case studies across three code families show that typed metadata makes code-specific decisions verifiable and retargetable: on the 20-qubit Pathfinder at Oak Ridge National Laboratory, decoding reduced logical readout error by 3.3 times. They also expose hidden architectural costs: for a three-qubit Quantum Fourier Transform (QFT-3), a magic-state factory requires five times the peak qubits of the encoded circuit. These experiences demonstrate a practical foundation for developing and evaluating fault-tolerant compiler transformations across software and hardware targets.
\end{abstract}

\keywords{fault-tolerant quantum computing, quantum error correction, compiler intermediate representation, MLIR, quantum compilation}

\author{Narasinga Rao Miniskar}
\authornote{These authors contributed equally to this research.}
\affiliation{%
  \institution{Oak Ridge National Laboratory}
  \city{Oak Ridge}
  \state{Tennessee}
  \country{USA}
}

\author{Elaine Wong}
\authornotemark[1]
\affiliation{%
  \institution{Oak Ridge National Laboratory}
  \city{Oak Ridge}
  \state{Tennessee}
  \country{USA}
}

\author{Seyong Lee}
\authornotemark[1]
\affiliation{%
  \institution{Oak Ridge National Laboratory}
  \city{Oak Ridge}
  \state{Tennessee}
  \country{USA}
}

\author{Vicente Leyton-Ortega}
\affiliation{%
  \institution{Oak Ridge National Laboratory}
  \city{Oak Ridge}
  \state{Tennessee}
  \country{USA}
}

\author{Jeffrey S. Vetter}
\affiliation{%
  \institution{Oak Ridge National Laboratory}
  \city{Oak Ridge}
  \state{Tennessee}
  \country{USA}
}

\author{Travis S. Humble}
\affiliation{%
  \institution{Oak Ridge National Laboratory}
  \city{Oak Ridge}
  \state{Tennessee}
  \country{USA}
}

\thanks{This manuscript has been authored by UT-Battelle, LLC, under Contract No.\ DE-AC05-00OR22725 with the U.S.\ Department of Energy. The United States Government retains and the publisher, by accepting the article for publication, acknowledges that the United States Government retains a non-exclusive, paid-up, irrevocable, world-wide license to publish or reproduce the published form of this manuscript, or allow others to do so, for the United States Government purposes. The Department of Energy will provide public access to these results of federally sponsored research in accordance with the DOE Public Access Plan.}

\maketitle


\section{Introduction}
\label{sec:intro}

The transition from noisy intermediate-scale quantum (NISQ) devices to fault-tolerant quantum computers depends on quantum error correction (QEC)~\cite{shor1995,steane1996}. A QEC scheme encodes each logical qubit in a set of physical qubits. This redundancy detects and corrects physical errors while preserving the logical state~\cite{gottesman1997}, but at substantial cost: practical fault-tolerant algorithms may require millions of physical qubits~\cite{fowler2012,gidney2021}.

The software ecosystem for fault-tolerant compilation remains less developed than its classical counterpart. Quantum compilers such as Qiskit~\cite{qiskit}, Cirq~\cite{cirq}, and tket~\cite{tket} primarily optimize gates for NISQ devices. MLIR-based quantum intermediate representations (IRs) include PennyLane/Catalyst~\cite{bergholm2018} and NVIDIA's CUDA-Q Quake dialect~\cite{cudaq,quakeDialect}. LLVM-level IRs and programming frameworks include the quantum intermediate representation (QIR)~\cite{qir2021}, ScaffCC~\cite{scaffcc2015}, and XACC~\cite{xacc2019}. These IRs and systems do not natively represent error correcting code (ECC) parameterized logical qubits, syndrome extraction, or magic-state primitives as first-class IR entities. Hopf et al.~\cite{hopf2026mlir} identify this lack of fault-tolerant abstractions in current MLIR stacks as an open problem.

Recent work addresses several parts of the fault-tolerant compilation stack. Lattice-surgery systems such as TopoLS~\cite{topols2026}, Pure Magic~\cite{litinski2025puremagic}, and movable-qubit routing~\cite{hopf2025movable} optimize space-time volume. The tableau-based quantum compiler (TQC)~\cite{wang2025tqc} reduces runtime through tableau-based methods, while early-FTQC compilers~\cite{sharma2026earlyftqc,quantem2025} target resource-constrained systems. None of these systems, however, provides a composable MLIR pass pipeline. Non-Clifford Fusion~\cite{noncliffordfusion2025} and lazy resynthesis~\cite{lazyresynth2025} reduce the cost of $T$ gates. Experiments in magic-state cultivation~\cite{gidney2025cultivation} and logical magic-state distillation~\cite{quera2025distillation} show progress toward practical non-Clifford resources. Real-time QEC decoders include AlphaQubit~2~\cite{alphaqubit2025}, Sparse Blossom~\cite{higgott2025sparseblossom}, and Riverlane's local clustering decoder~\cite{riverlane2025lcd}. These advances focus on algorithms, simulation, or hardware. A compiler IR for representing and transforming fault-tolerant abstractions is still needed.

This work addresses that representation gap. It does not propose a new error-correcting code or decoder. The central question is how a compiler can preserve code-specific information while lowering a logical algorithm to physical circuits, simulators, and hardware interfaces. MLIR types and lowering passes keep these decisions explicit throughout compilation.

The ECC-aware type system, the two-level lowering design, and the IR-resident resource analysis are new in this work. The implementation builds on established components. These include MLIR and Operation Definition Specification (ODS), QIR, Stim, PyMatching, standard error-correcting codes, and published distillation protocols. This separation defines the scope of the paper's novelty.

To that end, we present \texttt{FTQC}, a domain-specific MLIR dialect for compiling fault-tolerant quantum circuits~\cite{lattner2021mlir}. This paper makes the following contributions:

\begin{itemize}\itemsep2pt
  \item We introduce an ECC-aware type system for fault-tolerant programs. It represents logical qubits, syndrome registers, magic states, and Pauli frames as first-class SSA values. Each logical-qubit type records its code family and distance. The verifier uses this to reject invalid operations, such as gates between qubits in incompatible codes.

  \item We define a two-level lowering model that connects the ECC-typed logical dialect to a shared physical-qubit dialect. Different back ends therefore consume the same physical IR. We also implement a resource-analysis pass that derives physical-qubit counts and logical error rates from type metadata in the IR.

  \item We evaluate \texttt{FTQC} with three quantum error-correcting codes: the Steane, surface, and color codes. Our benchmarks cover Clifford memory, magic-state distillation, variational layers, lattice-surgery routing, and arbitrary-angle rotations synthesized into Clifford+$T$. The evaluation reports physical-qubit requirements, syndrome-extraction counts, and distillation overhead. In the peak-space model used here, one serially reused $T$-state factory dominates the 3-qubit Quantum Fourier Transform (QFT-3) qubit estimate.

  \item We provide an open-source, end-to-end implementation with OpenQASM~3.0~\cite{cross2022openqasm3}, QIR~\cite{qir2021}, and Stim~\cite{gidney2021stim}, and cross-check the logical results with Qiskit Aer~\cite{qiskit} and the QIR execution engine (QIR-EE). We evaluate noisy syndrome extraction with Stim simulations and PyMatching decoding~\cite{pymatching2021}. These experiments validate noisy-QEC behavior but do not demonstrate threshold scaling.
\end{itemize}


\section{Background}
\label{sec:background}

\texttt{FTQC} rests on three pillars: the quantum circuit model that defines the programs, the error-correcting codes that encode them fault-tolerantly, and the MLIR infrastructure that represents and transforms them.

\subsection{Quantum Circuits}
A quantum circuit is the standard model for quantum algorithms~\cite{nielsen2010}: a sequence of gates on qubits, with measurements that yield classical outcomes. It applies a unitary $U$ to an initial state, typically $\lvert 0 \rangle^{\otimes n}$. Fault-tolerant circuits also use mid-circuit measurement and classical feed-forward, both essential to syndrome extraction and correction.
Gates are reversible linear operations. Single-qubit gates include $\{X,Y,Z,H,S,T\}$; multi-qubit gates such as $\mathrm{CNOT}$ generate entanglement. The Clifford set $\{H,S,\mathrm{CNOT}\}$ is classically simulable (Gottesman-Knill~\cite{aaronson2004}); adding the non-Clifford $T$ gate gives a universal set that approximates any unitary to arbitrary accuracy~\cite{dawson2005}.
Circuits are often modeled as directed acyclic graphs, which expose parallelism and enable optimizations such as gate cancellation and depth reduction. Compilation flows lower high-level programs to circuits before hardware mapping, and increasingly build these stacks on classical infrastructure such as LLVM~\cite{lattner2004llvm} and MLIR~\cite{lattner2021mlir}.

\subsection{Quantum Error Correction}

On real hardware, physical errors corrupt the computation. Quantum error correction encodes each logical qubit into many physical qubits, which reshapes the circuit. Logical Clifford gates are usually transversal~\cite{gottesman1997}, whereas non-Clifford gates such as $T$ need magic-state distillation and injection~\cite{bravyi2005} with many ancilla qubits and stabilizer measurements. Circuit structure, especially $T$-count and $T$-depth, thus dominates resource cost in fault-tolerant quantum computing, and a corresponding compiler must track the error-correcting code and its parameters.

An $[\![n,k,d]\!]$ code encodes $k$ logical qubits into $n$ physical qubits with minimum \textit{distance} $d$, the smallest number of physical errors that can cause a logical error. After one correction round, the logical error rate scales as
\begin{equation}
  p_L \approx C \cdot \big( p/p_\text{th} \big)^{\!\lfloor d/2\rfloor+1},
  \label{eq:logerrorprob}
\end{equation}
where $p$ is the physical error rate and $p_\text{th}$ the fault-tolerance threshold. The constant $C$ is exact at low $p$ but in general depends on the code, noise model, and decoder. A code built from two compatible classical linear codes, with separate $X$- and $Z$-stabilizer checks, is a Calderbank-Shor-Steane (CSS) code~\cite{steane1996}. The dialect supports three CSS families, each trading off qubit overhead, threshold, and transversal-gate support differently.

The simplest is the {\bf Steane} \steane{} {\bf code}, encoding one logical qubit into seven physical qubits at distance three. It has transversal $\{X,Z,H,S,\mathrm{CNOT}\}$ gates but injects a magic state for the non-Clifford $T$~\cite{bravyi2005}. For higher distances, the rotated planar {\bf surface code}~\cite{fowler2012} scales: it uses $d^2+(d-1)^2$ physical qubits at distance $d$ and reaches $p_\text{th}\approx 1\%$, the highest threshold of any established topological code, though its Hadamard needs lattice surgery rather than a transversal gate. The triangular {\bf 2D color code}~\cite{bombin2006} makes the whole Clifford group transversal~\cite{landahl2011fault} at a lower threshold, using $(3d^2+1)/4$ data qubits at distance $d$.

In all three families the $T$ gate is not transversal and instead comes from {\bf magic-state distillation}~\cite{bravyi2005}. A 15-to-1 (or 7-to-1) protocol consumes 15 (or 7) noisy $|T\rangle$ copies to purify one. This is costly: one 15-to-1 factory on the Steane code needs $195$ physical qubits ($15\times7=105$ data qubits and $15\times6=90$ ancilla qubits for stabilizer measurements). Managing this overhead is a central task for a fault-tolerant compiler.

\subsection{LLVM and MLIR}

These code families, syndrome protocols, and distillation strategies need an infrastructure that expresses domain-specific abstractions while keeping the optimization power of a production compiler. LLVM~\cite{lattner2004llvm} supplies a typed SSA representation and a modular pass manager. MLIR~\cite{lattner2021mlir} adds a \emph{dialect-based} IR, where a dialect is a named namespace of custom types, attributes, and operations. Several dialects coexist in one module, enabling gradual lowering from high-level abstractions to machine instructions. In \texttt{FTQC}, ECC-parameterized logical-qubit types, physical-gate operations, and classical control flow share one module and lower independently.

MLIR supports this lowering with pattern rewriting (\texttt{RewritePatternSet}), type conversion (\texttt{TypeConverter}), and a verifier that checks operation type invariants at each stage. The Operation Definition Specification (ODS) framework synthesizes C++ classes, parsers, printers, and verifiers from TableGen, cutting the boilerplate of a new dialect such as \texttt{FTQC}. The pass manager runs transformation and analysis passes as a directed pipeline, with statistics recording, IR snapshot debugging, and concurrent execution.


\section{\texttt{FTQC} Dialect Design}
\label{sec:design}

This section introduces the main components and motivation of our \texttt{FTQC} dialect.

\subsection{Design Principles}

Three principles guide \texttt{FTQC}. \textbf{Type safety} places ECC parameters in type declarations. The compiler therefore rejects an operation that mixes a Steane qubit with a surface-code qubit. \textbf{SSA qubit semantics} represent logical qubits as immutable SSA values. Each gate consumes its inputs and produces new values. This form supports standard data-flow analysis without pointer aliasing. \textbf{Progressive lowering} places \texttt{FTQC} operations at the top of a staged hierarchy. Successive passes lower them to physical hardware instructions. These principles keep code-specific decisions visible during transformation and verification.

\addfigcols{fig:ftqc-design}{The two-level  compiler pipeline.  \textbf{Level~1} (logical IR, \texttt{ftqc.*}) applies the transformational and optimization passes (encoding, syndrome insertion, transversal decomposition, magic-state distillation, and Pauli-frame optimization) alongside the analysis passes (fault-path, threshold, resource estimation).  The \texttt{--ftqc-expand-physical} pass bridges to \textbf{Level~2} (physical IR, \texttt{ftqc.qphys.*}), where an optional hardware layout is assigned and every back end (OpenQASM~3, QIR, Stim, and IQM~JSON) lowers the \emph{one} shared physical IR.}{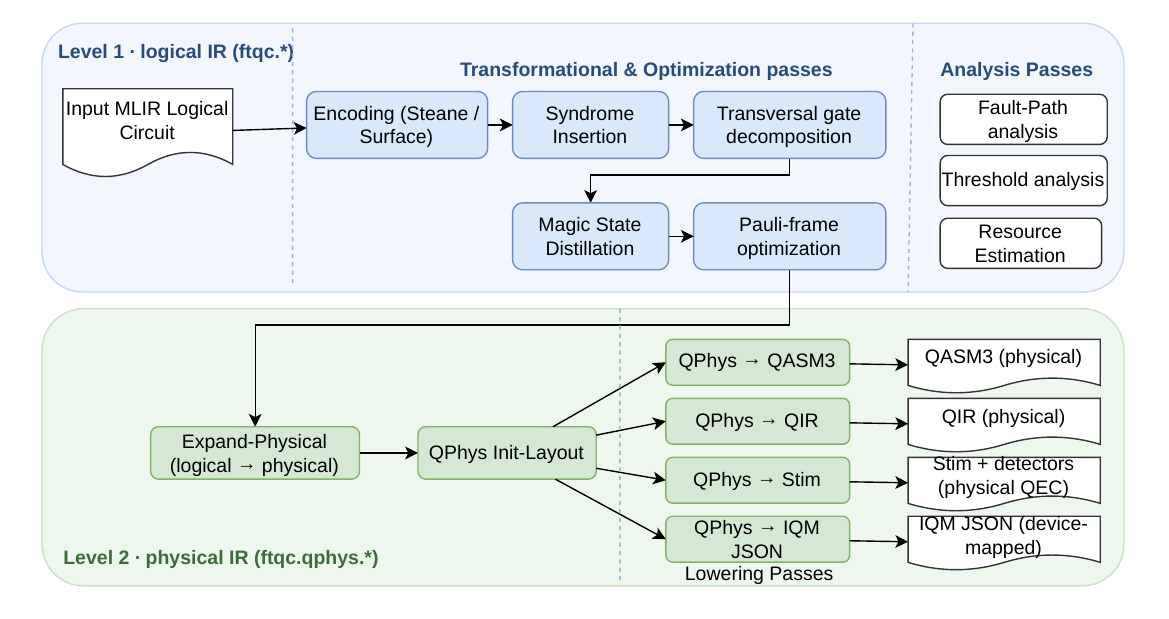}

For example, \texttt{logical\_cnot} requires compatible ECC attributes on both operands. A CNOT between a Steane qubit and a surface-code qubit fails verification before lowering. An untyped gate-level IR would need a separate analysis to find the same mismatch.

\subsection{Two-Level Dialect Structure}
\label{sec:twolevel}

FTQC has two dialect levels (Table~\ref{tab:levels}) joined by one pass. Every back end consumes the same physical IR. It does not derive a separate physical layout. \textbf{Level~1} (L1) is the logical dialect (\texttt{ftqc.*}). It represents error-corrected qubits and stores the ECC family and distance in the type (\texttt{\#ftqc.ecc<K,d>}). \textbf{Level~2} (L2) is the physical dialect (\texttt{ftqc.qphys.*}). Each value represents one hardware qubit with standard quantum SSA semantics.

The \texttt{--ftqc-expand-physical} pass is the only bridge from Level~1 to Level~2. It expands each \texttt{!ftqc.logical\_qubit} into a block of \texttt{!ftqc.qphys\_qubit} values. For the Steane \steane{} code, the block contains seven data qubits and three shared ancilla. The ancilla are reused for X- and Z-stabilizer extraction. The pass also rewrites logical gates and syndromes as explicit physical circuits. All four back ends consume this common IR after an optional hardware layout step. Routing and device mapping remain within Level~2 through operations such as \texttt{qphys.swap}. The Stim back end rebuilds \texttt{DETECTOR} and logical \texttt{OBSERVABLE\_INCLUDE} records from QEC metadata. The metadata records the role, stabilizer, and round of each measurement. Stim therefore evaluates the same physical circuit represented for the other back ends.

\begin{table}[h]
\centering\small
\caption{Separation of the two \texttt{FTQC} dialect levels. L1 ops are \texttt{ftqc.*}, L2 ops are \texttt{ftqc.qphys.*}, and pass names use the \texttt{--ftqc-} prefix. Every back end lowers the shared physical L2 IR produced by \texttt{expand-physical}.}
\label{tab:levels}
\vspace{-4pt}
\resizebox{\columnwidth}{!}{%
\begin{tabular}{@{}p{1.1cm}p{2.6cm}p{3.3cm}p{2.2cm}@{}}
\toprule
\textbf{Level} & \textbf{Values / types} & \textbf{Representative ops} & \textbf{Lowered by} \\
\midrule
Logical & \texttt{logical\_qubit}, \texttt{syndrome}, \texttt{magic\_state}, \texttt{pauli\_frame} & \texttt{init\_zero}, \texttt{logical\_\{h, cnot, t\}}, \texttt{steane\_\{x, z\}\_syndrome}, \texttt{correct}, \texttt{distill\_15to1}, \texttt{lattice\_surgery} & \texttt{expand-} \texttt{physical} \\
\addlinespace[2pt]
Physical & \texttt{qphys\_qubit} (one hardware qubit) & \texttt{qphys.alloc}, \texttt{qphys.\{h, x, z, s, t, cx, cz, swap\}}, \texttt{qphys.measure} & \texttt{qphys-to-} \texttt{qasm3}, \texttt{qphys-to-} \texttt{iqm-json} \\
\bottomrule
\end{tabular}%
}
\end{table}

\subsection{Type System}

\texttt{FTQC} introduces five types (Table~\ref{tab:types}), defined and validated through MLIR’s ODS TableGen \texttt{TypeDef} construct. The \texttt{\#ftqc.ecc<K,d>} attribute, with $K\in\{\texttt{steane},\texttt{surface},\texttt{color\_code}\}$ and integer $d$, is embedded in the logical-qubit type.

\begin{table}[h]
\centering\footnotesize
\caption{FTQC custom types.}
\label{tab:types}
\vspace{-4pt}
\begin{tabular}{@{}lp{2.2cm}@{}}
\toprule
\textbf{Type} & \textbf{Meaning} \\
\midrule
\texttt{!ftqc.logical\_qubit}\ \texttt{<\#ftqc.ecc<K,d>>} & Logical qubit \\
\texttt{!ftqc.syndrome<N>}      & $N$-bit syndrome \\
\texttt{!ftqc.magic\_state}     & $|T\rangle$ magic state \\
\texttt{!ftqc.magic\_state<CCZ>}& $|CCZ\rangle$ state \\
\texttt{!ftqc.pauli\_frame<N>}  & $2N$-bit Pauli frame \\
\bottomrule
\end{tabular}
\end{table}

\vspace{-2pt}
\subsection{Operation Set}

\texttt{FTQC} defines 31 operations (ops) in 7 categories (Table~\ref{tab:ops}).

\textbf{Qubit lifecycle.}
\texttt{ftqc.init\_zero}, \texttt{ftqc.init\_plus}, and \texttt{ftqc.dealloc} allocate and free logical qubits. \texttt{ftqc.encode} and \texttt{ftqc.decode} mark the transition between physical and logical representations, letting passes track encoding overhead.

\textbf{Clifford gates.}
The single-qubit Clifford gates (\texttt{logical\_x}, \texttt{logical\_y}, \texttt{logical\_z}, \texttt{logical\_h}, \texttt{logical\_s}) carry the \texttt{SameOperandsAndResultType} trait, so their result type is inferred. The two-qubit gates \texttt{logical\_cnot} and \texttt{logical\_cz} return two output values.

\textbf{Non‑Clifford gates.}
Magic-state injection splits into preparation and use: \texttt{prepare\_magic\_t} produces a $|T\rangle$ resource and \texttt{prepare\_magic\_ccz} a $|CCZ\rangle$ state, which the injection ops \texttt{logical\_t} and \texttt{logical\_toffoli} consume. Toffoli uses a Clifford+T decomposition for three-qubit controlled operations.

\textbf{Syndrome extraction.}
Five ops cover syndrome extraction: the Steane X/Z pair (\texttt{steane\_x\_syndrome}, \texttt{steane\_z\_syndrome}), the flag fault-tolerant syndrome (\texttt{flag\_syndrome}), the Shor cat-state syndrome (\texttt{shor\_syndrome}), and a generic \texttt{extract\_syndrome}. Each returns the post-extraction logical qubit and a classical syndrome value, preserving SSA data-flow.

\textbf{Error correction and measurement.}
\texttt{ftqc.correct} applies a Pauli correction derived from a syndrome, \texttt{ftqc.logical\_measure} measures a logical qubit to a classical \texttt{i1} result, and \texttt{ftqc.ec\_round} wraps a complete syndrome-extraction and correction round. A Pauli frame is an alternative to physical correction: it records Pauli errors classically and folds them into the interpretation of later operations and measurements, avoiding many physical corrections during Clifford computation.

\textbf{Pauli‑frame management.}
\texttt{ftqc.create\_frame} allocates a $2N$-bit Pauli frame, \texttt{ftqc.pauli\_frame\_update} defers a correction to a classical register, and \texttt{ftqc.apply\_frame} applies the accumulated byproduct corrections before final measurement.

\textbf{Distillation and teleportation.}
\texttt{ftqc.distill\_15to1} and \texttt{ftqc.distill\_7to1} model the two canonical magic-state distillation protocols, and \texttt{ftqc.gate\_teleport} encodes gate teleportation from a pre-prepared resource state.

\begin{table}[h]
\centering\small
\caption{\texttt{FTQC} operation categories and counts.}
\label{tab:ops}
\vspace{-4pt}
\begin{tabular}{@{}lc@{}}
\toprule
\textbf{Category} & \textbf{Ops} \\
\midrule
Qubit lifecycle (init, encode, decode, dealloc) & 5 \\
Clifford gates (single- and two-qubit)          & 7 \\
Non-Clifford gates (T, Toffoli, magic prep)     & 4 \\
Syndrome extraction                             & 5 \\
Error correction \& measurement                 & 4 \\
Pauli-frame management                          & 3 \\
Distillation \& teleportation                   & 3 \\
\midrule
\textbf{Total}                                  & \textbf{31} \\
\bottomrule
\end{tabular}
\end{table}

Listing~\ref{lst:bell} shows the Bell-state circuit in the \texttt{FTQC} IR.

\begin{lstlisting}[caption={An example of a circuit expressed in \texttt{FTQC} IR.},
                   label=lst:bell, float=t]
func.func @bell_state_ft() -> (i1, i1) {
  %lq0 = ftqc.init_zero : <<steane, 3>>
  %lq1 = ftqc.init_zero : <<steane, 3>>
  %h0  = ftqc.logical_h %lq0 : <<steane, 3>>
  %c, %t = ftqc.logical_cnot %h0, %lq1
      : (<<steane, 3>>, <<steane, 3>>)
      -> (<<steane, 3>>, <<steane, 3>>)
  %b0 = ftqc.logical_measure %c
      : (<<steane, 3>>) -> i1
  %b1 = ftqc.logical_measure %t
      : (<<steane, 3>>) -> i1
  func.return %b0, %b1 : i1, i1
}
\end{lstlisting}


\section{Transformation Passes}
\label{sec:passes}

\texttt{FTQC} provides 18 passes on MLIR's pass infrastructure.
Three composite passes bundle common sequences: \texttt{--ftqc-steane-pipeline}, \texttt{--ftqc-surface-pipeline}, and \texttt{--ftqc-distill-pipeline}. The other 15 cover logical encoding, analysis, decomposition, and lowering to three hardware back ends.

Every pass consumes and produces valid MLIR, so the verifier checks type and operation invariants between stages, and intermediate modules can be inspected or individual stages swapped. The next subsections describe the main analyses and transformations.

\subsection{Syndrome Insertion}
The \texttt{--ftqc-insert-syndrome} pass makes one forward traversal and inserts a \texttt{steane\_x\_syndrome}/\texttt{steane\_z\_syndrome} pair after each logical gate. The syndrome width $N$ comes from the ECC attribute: $N=6$ for Steane at $d=3$, and $N=24$ for the rotated planar surface code at $d=5$ ($d^2-1$; Table~\ref{tab:ecc}).

\subsection{Transversal Gate Decomposition}
For the Steane \steane{} code, the set of transversal Clifford operations consists of $\{X, Z, H, S, \text{CNOT}\}$.
In the Steane code $S$ is also transversal; $T$ is not, so \texttt{FTQC} implements it by magic-state injection and gate teleportation. In the surface code neither $S$ nor $T$ is transversal, which requires lattice surgery or magic-state injection, and lattice surgery is preferred to implement $CNOT$.

\subsection{Resource Estimation}
The \texttt{--ftqc-resource-estimate} pass collects six metrics in one traversal. It infers the logical-qubit count from \texttt{init\_zero}, \texttt{init\_plus}, and the argument types of \texttt{func::FuncOp}; the last covers circuits like QFT-3 that receive qubits as function arguments. It then counts $T$ gates, CNOT gates, other Clifford gates, and individual stabilizer-type syndrome extractions, taking one X-type and one Z-type extraction after each logical gate as one complete correction cycle. Finally it computes the peak distillation-qubit count: one serially reused 15-to-1 factory with the Steane CSS code contributes 195 qubits, independent of $n_T$.

The pass derives physical-qubit requirements and logical error rates from the ECC metadata in the IR (Table~\ref{tab:ecc}). A Steane logical qubit uses $7+6$ physical qubits. A rotated planar surface-code qubit of distance $d$ uses $d^2+(d^2-1)$ qubits. A 2D color-code qubit at odd distance $d$ uses $\frac{3d^2+1}{4}+\frac{d^2-1}{2}$ qubits. The pass approximates the logical error rate with \eqref{eq:logerrorprob},
where $p = 10^{-3}$ denotes the typical superconducting gate error rate and $p_\text{th}$ is taken from Table~\ref{tab:ecc}. 
Accumulated over all extractions, the coarse leading-order estimate (not a rigorous bound) is $p_{\rm total} = p_L \times \text{SE} \times n_L$, with SE the number of stabilizer-type syndrome extractions. It comes straight from the IR and differs from the \emph{per-shot circuit-level failure probability} $p_{\rm shot}$ that the Stim back end measures by simulating a physical syndrome-extraction circuit under depolarizing noise and decoding with MWPM (Section~\ref{sec:stim}). For Steane $d=3$ at $p=10^{-3}$, \eqref{eq:logerrorprob} gives $p_L=10^{-3}$: Bell (four extractions, two logical qubits) gives $p_{\rm total}\approx8\times10^{-3}$, and QFT-3 (36 extractions, three logical qubits) gives $p_{\rm total}\approx0.11$. The pass stores the total physical-qubit count and estimated logical error in an \texttt{ftqc.resource\_annotation} operation.

\begin{table}[h]
\centering\small
\caption{ECC physical parameters employed by the resource estimator. $d$: code distance; $p_\text{th}$: fault‑tolerance threshold, where surface code refers to rotated planar surface code, and 2D color code refers to standard triangular 2D color code.}
\label{tab:ecc}
\vspace{-4pt}
\begin{tabular}{@{}lccc@{}}
\toprule
\textbf{Code} & \textbf{Data qubits} & \textbf{Ancilla} & $p_\text{th}$ \\
\midrule
Steane $[\![7,1,3]\!]$   & 7          & 6             & 1\% \\
Surface code, dist.\ $d$ & $d^2$      & $d^2-1$       & 1\% \\
2D color code, dist.\ $d$& $(3d^2+1)/4$ & $(d^2-1)/2$   & 0.3\% \\
\bottomrule
\end{tabular}
\end{table}

\subsection{Pauli‑frame Optimization}
The \texttt{--ftqc-pauli-frame-opt} pass replaces each Pauli \texttt{ftqc.correct} with \texttt{ftqc.pauli\_frame\_update}, deferring all Pauli corrections to one final classical post-processing phase and cutting the number of physical corrections.

\subsection{OpenQASM 3.0 Emission}
The \texttt{--ftqc-lower-to-qasm3} back end~\cite{cross2022openqasm3} converts a post-pipeline FTQC module into an OpenQASM~3.0 for logical validation. It maps each logical qubit and gate to one QASM qubit and gate. It does not expand physical qubits or encode syndrome circuits. Simulation therefore checks the algorithm and the emitter. The Stim back end evaluates the noisy syndrome path described in Section~\ref{sec:stim}.

\begin{itemize}\itemsep0pt
  \item \texttt{init\_zero} / \texttt{init\_plus} $\to$ \texttt{qubit q$_i$;}
        (\texttt{h q$_i$;} for $|+\rangle$).
  \item Single-qubit Clifford gates $\to$ \texttt{h}, \texttt{x},
        \texttt{y}, \texttt{z}, \texttt{s}.
  \item \texttt{logical\_t} $\to$ \texttt{t q$_i$;} (magic-state value
        consumed silently; no qubit allocation emitted).
  \item \texttt{logical\_toffoli} $\to$ \texttt{ccx q$_a$, q$_b$, q$_c$;}.
  \item \texttt{logical\_measure} $\to$ \texttt{bit b$_j$ = measure q$_i$;}.
  \item Syndrome, correction, and Pauli-frame ops $\to$ QASM comments (default \texttt{--include-comments=true}). QASM3 has no native error-correction construct, so this pass targets logical-level validation only.
\end{itemize}

The emitted string begins with \texttt{OPENQASM 3.0;} and \texttt{include ``stdgates.inc'';} so that the Qiskit QASM3 parser recognizes the standard gate set.

\subsection{QIR Emission}

The \texttt{--ftqc-lower-to-qir} pass translates a post-pipeline \texttt{FTQC} module to QIR~\cite{qir2021} Base Profile LLVM IR, expanding every logical qubit to physical qubits. A Steane~\cite{steane1996} logical qubit becomes 7 data qubits with a fixed encoder and transversal gate mappings; a surface-code qubit of distance~$d$ becomes a $d \times d$ physical grid. The pass maps (i) initialization opcodes to their encoding circuits; (ii) logical single- and multi-qubit gates to transversal implementations across the logical block; and (iii) measurement to \texttt{mz} on every physical qubit, leaving the XOR post-processing to the caller. 
The Base Profile does not support the mid-circuit measurement and feed-forward required by executable syndrome correction. The emitter therefore records syndrome, correction, and Pauli-frame operations as comments rather than executable QIR. It does not emit a physical magic-state injection circuit. 
The \texttt{--ftqc-lower-to-qir-ns} path omits syndromes and reproduces the logical results via QIR-EE/Aer (Table~\ref{tab:simresults}); with syndromes inserted the pass supports the full decoding pipeline.

\subsection{Stim Stabilizer Circuit Emission}
\label{sec:stim}
The \texttt{--ftqc-lower-to-stim} pass converts a post-pipeline \texttt{FTQC} module into a Stim~\cite{gidney2021stim} stabilizer circuit for Monte-Carlo shot-noise simulation and minimum weight perfect matching (MWPM) decoding. For each logical Steane qubit it allocates a 13-qubit block (7 data, 3 X-ancilla, 3 Z-ancilla) and emits (i) encoding via \texttt{R} and \texttt{H} on check qubits \{0, 1, 3\} then a CNOT cascade; (ii) syndrome rounds of ancilla reset, Hadamard, CNOT fan-out, measurement, and \texttt{DETECTOR} records comparing current and previous syndrome bits; (iii) transversal logical gates as one- and two-qubit Clifford instructions; and (iv) final data-qubit measurement tagged with \texttt{OBSERVABLE\_INCLUDE}. Because Stim is Clifford-only, a logical $T$ becomes a transversal $S$, a fixed Clifford stand-in that preserves stabilizer error propagation but not the non-Clifford action, and magic-state injection is not represented; the Stim results thus characterize syndrome extraction and decoding along the circuit's Clifford structure, not the fidelity of the non-Clifford computation. Depolarizing noise enters at rate $p$ via \texttt{DEPOLARIZE1} after one-qubit gates, \texttt{DEPOLARIZE2} after two-qubit gates, and \texttt{X\_ERROR} before measurements.

\subsection{Multi-controlled and arbitrary-rotation synthesis}
\label{sec:synth}
Two front-end passes lower gates outside the native \{Clifford,\,$T$,\,Toffoli\} set, so the IR can ingest oracle and rotation structure.  \texttt{--ftqc-decompose-mcx} rewrites an $n$-controlled \texttt{logical\_mcx} into a Toffoli network: $0/1/2$ controls map to $X$/CNOT/Toffoli, while $n\ge3$ uses the Barenco clean-ancilla V-chain with $n{-}2$ ancilla and $2(n{-}1){-}1$ Toffoli gates that are uncomputed so the ancilla return to $|0\rangle$.  \texttt{--ftqc-synthesize-rz} approximates a logical $R_z(\theta)$ (\texttt{logical\_rz}) by a Clifford+T word using the Solovay-Kitaev algorithm over the generator set $\{H,S,T,X,Y,Z,S^\dagger,T^\dagger\}$: dyadic multiples of $\pi/4$ are emitted exactly, and arbitrary angles are approximated to a precision fixed by the recursion depth (depth~2 attains $\varepsilon\approx5\times10^{-2}$; depth~3, $\approx2\times10^{-2}$), with the resulting T-count recorded as an attribute.  Both are semantics-checked by recomputing the emitted circuit's reversible (multi-CX) or unitary (Rz) action on small instances.


\section{Experimental Results}
\label{sec:experiments}

We evaluate on a node at Oak Ridge National Laboratory (ORNL): Ubuntu 22.04, AMD EPYC 7513 processor, 128 threads, 1 TB RAM. \texttt{FTQC} is built with LLVM/MLIR 22.0.0git at \texttt{-O2}. For stable numbers on this shared node, every compile-time measurement pins the process to one core (\texttt{taskset -c 0}) with MLIR threading disabled (\texttt{-{}-mlir-disable-threading}). Thread count has negligible effect, since these single-function circuits offer no inter-op parallelism. Wall-clock time is thus single-threaded and dominated by one-time library startup, so we isolate the in-pipeline pass cost with \texttt{-mlir-timing} (Fig.~\ref{fig:passtime}).

\subsection{Benchmark Circuits}
\label{sec:qft3}
\begin{itemize}
  \item \textbf{Bell state}: prepares \((|00\rangle+|11\rangle)/\sqrt{2}\) with a Hadamard then a CNOT, on two distance-3 Steane logical qubits; Clifford-only, no \(T\) gates.
  \item \textbf{GHZ‑4}: the four‑qubit Greenberger-Horne-Zeilinger (GHZ)
  state \((|0000\rangle+|1111\rangle)/\sqrt{2}\) is generated by one Hadamard gate and a chain of three CNOTs. The four logical qubits use the distance-3 Steane code. This example tests syndrome insertion as the CNOT depth increases.
  \item \textbf{Grover‑2}: a two-qubit search implementing \(H^{\otimes 2}\to\text{CZ}\to H^{\otimes 2}\cdot (X^{\otimes 2}\cdot\text{CZ}\cdot X^{\otimes 2})\cdot H^{\otimes 2}\), on a distance-5 surface code to exercise a non-Steane ECC. Its native two-qubit CZ oracle marks $|11\rangle$, so it is Clifford-only with no \(T\) gates.

  \item \textbf{QFT‑3}: a three-qubit QFT in a Clifford+\(T\) decomposition with six \(T\) gates and nine CNOTs; the rest are Clifford. The Steane pipeline implements $S$ transversally rather than as $T\!\cdot\!T$, so QFT-3 is the only main benchmark exercising distillation-factory allocation.
\end{itemize}

Two \emph{parametrized} families scale with the logical-qubit count $n$, so the resource estimates (Table~\ref{tab:resources}) and the scaling and per-pass figures (Figs.~\ref{fig:compile_time},~\ref{fig:passtime}) span a range of $n$. Both use the Steane $[\![7,1,3]\!]$ code.
\begin{itemize}
  \item \textbf{QAOA} (ring): one Quantum Approximate Optimization Algorithm (QAOA) cost-plus-mixer layer on an $n$-node ring at a fixed Clifford+$T$ angle. The cost layer applies $R_{zz}(\pi/4)=\text{CNOT}\cdot T\cdot\text{CNOT}$ on each ring edge and the mixer a $R_z(\pi/4)\!\approx\!T$ per qubit, so gate and $T$ counts grow linearly in $n$.
  \item \textbf{VQE} (hardware-efficient ansatz): a Variational Quantum Eigensolver (VQE) that employs two single-qubit rotation sublayers (a fixed $R_z(3\pi/4)=S\cdot T$ per qubit) around a linear CNOT entangling ladder, the canonical hardware-efficient ansatz (HEA) shape; $T$-count grows linearly in $n$.
\end{itemize}

\subsection{Resource Estimation}
Table~\ref{tab:resources} reports the \texttt{FTQC} pipeline's resource estimates; each row is the \texttt{--ftqc-resource-estimate} output.

\begin{table}[h]
\centering\small
\caption{\texttt{FTQC} resource estimates.  $n_L$: logical qubits; $T$: $T$-gate count;
CNOT: CNOT count; Cl: other Clifford gates; SE: stabilizer-type syndrome extractions;
Dist.~$q$: peak distillation qubits (QFT-3: 195 = 105 data + 90 ancilla for one serially reused 15-to-1 Steane factory); $n_\text{phys}$: total physical qubits (data + ancilla + distillation); $p_{\rm total}$: coarse leading-order logical-error \emph{estimate} (not a rigorous bound) from \eqref{eq:logerrorprob}, accumulated over SE at $p=10^{-3}$. Code abbreviations: St.\ = Steane $[\![7,1,3]\!]$; Su.\ = rotated surface code. The last two rows are the parametrized families at $n=8$; these are the same instances plotted in Fig.~\ref{fig:stim_qec}.}
\label{tab:resources}
\vspace{-4pt}
\resizebox{\columnwidth}{!}{%
\begin{tabular}{@{}lrrrrrrrc@{}}
\toprule
\textbf{Circuit} & $n_L$ & $T$ & CNOT & Cl & SE & Dist. q & $n_\text{phys}$ & $p_\text{total}$ \\
\midrule
Bell (St.$d$3)   & 2 & 0 & 1  &  1 &  4 &   0 &  26 & $8{\times}10^{-3}$ \\
GHZ-4 (St.$d$3) & 4 & 0 & 3  &  1 &  8 &   0 &  52 & $3.2{\times}10^{-2}$ \\
Grover (Su.$d$5) & 2 & 0 & 0  & 12 & 20 &   0 &  98 & $4{\times}10^{-3}$ \\
QFT-3 (St.$d$3)  & 3 & 6 & 9  &  3 & 36 & 195 & 234 & $1.1{\times}10^{-1}$ \\
\midrule
QAOA ring (St.$d$3)  &  8 &  16 & 16 &  0 &  64 & 195 & 299 & $5.1{\times}10^{-1}$ \\
VQE HEA (St.$d$3)    &  8 &  16 &  7 & 16 &  46 & 195 & 299 & $3.7{\times}10^{-1}$ \\
\bottomrule
\end{tabular}%
}
\end{table}

\textbf{Syndrome-extraction accounting.} The SE column counts individual stabilizer-type extractions, not complete error-correction rounds. The default Steane method emits one $X$-type and one $Z$-type extraction per logical gate, so an $X{+}Z$ cycle contributes two; the \texttt{shor} and \texttt{flag} methods emit a combined extraction and contribute one. This matches the \texttt{--ftqc-resource-estimate} output: Bell tallies 4, GHZ-4 8, and QFT-3 36. The Stim experiment below inserts three cycles per gate, so its count is not comparable with SE.

\textbf{Distillation factory.}  QFT-3's six \(T\) gates share a \emph{single} 15-to-1 factory, charged once as a peak footprint of $15\times(7\text{ data}+6\text{ ancilla})=195$ physical qubits, not a per-\(T\) allocation. The model reuses one factory serially across all \(T\) states and reports peak space, not a space-time (qubit$\times$cycle) product; it omits throughput, latency, input-state preparation, and failure-and-retry.  Even so, the factory is five times the 39-qubit encoded circuit, which motivates \(T\)-count reduction; a space-time model is future work.  Grover's distance-5 surface code produces 24-bit syndrome vectors and the distance-3 Steane code 6-bit; the type system propagates these widths and infers physical-qubit counts from the \texttt{\#ftqc.ecc<K,d>} parameters without user annotation.

\textbf{Logical-error estimate.} Equation~\eqref{eq:logerrorprob} is a \emph{coarse leading-order heuristic}, not a rigorous bound: $p_L=C\,(p/p_\mathrm{th})^{\lfloor d/2\rfloor+1}$ with $C=0.1$ and a per-family threshold ($p_\mathrm{th}=10^{-2}$ for Steane and surface), at a fixed $p=10^{-3}$. These defaults are not fitted to any noise model or decoder, and the expression ignores circuit geometry, gate-location counts, syndrome-circuit design, measurement and reset error, decoder behavior, correlated faults, and factory failure. Accumulated linearly as $p_{\rm total}=p_L\times\text{SE}\times n_L$ (so $p_L=10^{-3}$ per extraction and logical qubit for Steane $d{=}3$), it gives $p_{\rm total}\approx8\times10^{-3}$ for Bell and $0.11$ for QFT-3 (Table~\ref{tab:resources}). We use it only for order-of-magnitude sizing; the decoded Stim sweep (Fig.~\ref{fig:stim_qec}) reports measured behavior.

\subsection{Simulation-based Validation}
We validate at four levels and do not conflate them.  \textbf{(1) Syntactic:} the QASM3 and QIR emitters produce parsable, executable output.  \textbf{(2) Logical-functional:} the Qiskit Aer and qsim experiments on QASM3/QIR confirm the IR encodes the intended algorithm, i.e., the expected output distributions.  Because the QASM3 path lowers syndrome, correction, and Pauli-frame operations to comments, L1 and L2 do \emph{not} exercise error-correction semantics.  \textbf{(3) Physical-encoding:} the Steane physical-circuit simulation checks the seven-qubit encoder and transversal gates at the qubit level.  \textbf{(4) Noisy-QEC:} the Stim experiment inserts physical syndrome rounds, applies MWPM decoding, and measures decoded failure rates under depolarizing noise.  We stop short of a fifth, \emph{fault-tolerance} level: no experiment shows threshold or sub-threshold suppression with distance scaling, and the QFT-3 Stim circuit is a Clifford stand-in; establishing fault tolerance is future work.

\textbf{Qiskit Aer simulation.}
For each benchmark, \texttt{ftqc-opt} emits OpenQASM~3 (\texttt{--ftqc-lower-to-qasm3}) and QIR (\texttt{--ftqc-lower-to-qir}). Qiskit parses the QASM and runs it on \texttt{AerSimulator}~\cite{qiskit} with 4{,}096 shots (8{,}192 for QFT‑3); the QIR runs through QIR-EE~\cite{wong2025qiree, qireerepo} on the same back end. Table~\ref{tab:simresults} summarizes the outcomes: Bell and GHZ‑4 show the expected 50/50 superposition, Grover outputs $|11\rangle$ deterministically, and QFT‑3 matches a state‑vector ground truth from the same OpenQASM~3. All four pass within five percentage points of the expected frequencies.

\begin{table}[h]
\centering\small
\caption{Simulations using Qiskit Aer for QASM3 and QIR-EE~\cite{wong2025qiree, qireerepo} for QIR with benchmark circuits (4096 shots unless noted).
A PASS result implies outcomes within five percentage points of the expected outcome.}
\label{tab:simresults}
\vspace{-4pt}
\begin{tabular}{@{}lll@{}}
\toprule
\textbf{Circuit} & \textbf{Expected} & \textbf{Result} \\
\midrule
Bell state     & 50\% $|00\rangle$, 50\% $|11\rangle$  & PASS \\
GHZ-4          & 50\% $|0000\rangle$, 50\% $|1111\rangle$ & PASS \\
Grover-2       & 100\% $|11\rangle$                    & PASS \\
QFT-3 (8192 shots) & Statevector-derived                & PASS \\
\bottomrule
\end{tabular}
\end{table}

\textbf{Qsim Cirq simulation.}
We further validated Grover‑2 and QFT‑3 with \texttt{qsimcirq}~\cite{villalonga2020qsim} (v0.22.0, Cirq 1.5.0), mapping the \texttt{FTQC} IR to Cirq primitives through the shared MLIR representation. Grover‑2 returned $|11\rangle$ with unit probability, and the QFT‑3 amplitudes matched the Qiskit state‑vector result once bit orderings are reconciled (Cirq most significant bit (MSB)‑first vs.\ Qiskit least significant bit (LSB)‑first, so $|100\rangle_{\text{Cirq}}=|001\rangle_{\text{Qiskit}}$). Both passed.

\textbf{Steane physical circuit simulation.}
A physical-level Steane encoding of Bell (14 qubits) and GHZ-4 (28 qubits), with $|0_L\rangle$ prepared by Hadamards on qubits 0,1,3 then nine CNOTs, transversal logical gates, and parity decoding of the seven data measurements, recovered the expected logical correlations on Qiskit's statevector (Bell) and stabilizer (GHZ-4) back ends.

\textbf{Stim stabilizer QEC simulation.}
We evaluate \texttt{--ftqc-lower-to-stim} on the four fixed benchmarks as Steane \steane{}-encoded round-trip circuits (prepare, apply, invert, measure), so the noiseless outcome is always $|0\rangle^{\otimes n}$ and Stim observables are deterministic. All four use the Steane $[\![7,1,3]\!]$ code for uniform comparison; the resource table independently uses the surface code for Grover to stress a different ECC family. The script inserts three syndrome rounds per logical gate, one more than the estimator's two-round (X+Z) model, sweeps the depolarizing rate $p$ over 18 log-uniform points from $10^{-4}$ to $1.5\times10^{-2}$, and decodes with PyMatching~\cite{dennis2002} MWPM. Structural validation confirms Bell (26 physical qubits, 30 detectors, 2 observables), GHZ (52 qubits, 66 detectors, 4 observables), Grover (26 qubits, 96 detectors, 2 observables), and QFT (39 qubits, 162 detectors, 0 observables). The QFT kernel has no final measurement (zero observables); the swept rate below uses the round-trip form, which carries a logical-$Z$ observable per qubit. Because Stim is Clifford-only, that round trip runs a \emph{Clifford stand-in} for QFT-3 (each $T$ replaced by $S$), so its rate reflects syndrome extraction and decoding along a QFT-shaped schedule, not the fidelity of the non-Clifford transform.

\begin{figure}[t]
\centering
\includegraphics[width=\columnwidth]{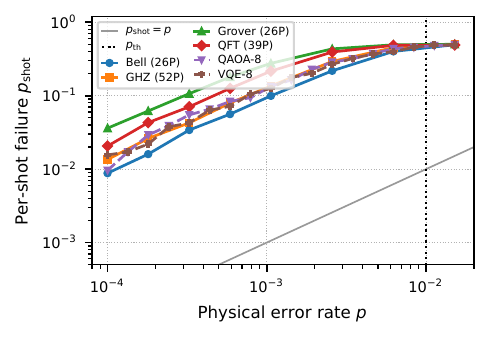}
\vspace{-0.15in}
\caption{\texttt{FTQC}+Stim per-shot circuit failure rate $p_{\rm shot}$ with MWPM decoding (PyMatching) for the six benchmarks (Steane $[\![7,1,3]\!]$). The four fixed circuits use 5{,}000 shots per point; the two parametrized families (QAOA, VQE at $n=8$, 104 physical qubits each) use 2{,}500 shots and are Clifford stand-ins with the non-Clifford angles abstracted (as for QFT-3). The gray diagonal marks $p_{\rm shot} = p$ (break-even); the dotted vertical line marks $p_\mathrm{th} \approx 1\%$.}
\label{fig:stim_qec}
\end{figure}

At $p = 10^{-3}$, the MWPM-decoded \emph{per-shot circuit failure rates} $p_{\rm shot}$ are 9.9\% for Bell, 13.6\% for GHZ, 27.6\% for Grover, and 21.6\% for QFT. The QAOA-8 and VQE-8 rates are 13.4\% and 12.6\%, respectively. Each of these two circuits uses 104 physical qubits. These values are cumulative round-trip failure probabilities over all logical gates. They are not per-qubit-per-round rates. They therefore differ from the analytic $p_{\rm total}$ in Table~\ref{tab:resources}, which applies \eqref{eq:logerrorprob} to the syndrome-extraction tally. The curves rise monotonically with $p$ and approach $0.5$. This trend is a sanity check, not a validation of decoding accuracy. Figure~\ref{fig:stim_qec} presents all six benchmarks.

\begin{figure}[t]
\centering
\includegraphics[width=\columnwidth]{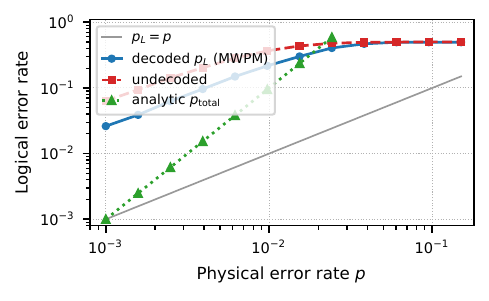}
\vspace{-1mm}
\caption{\texttt{FTQC}$+$Stim$+$MWPM decoded logical error rate for the Steane \steane{}
memory experiment (three syndrome rounds, $60{,}000$ shots/point) versus
physical error rate $p$: MWPM decoding (blue) lies below the undecoded rate
(red) at all $p$. The coarse heuristic $p_{\rm total}$ is shown in green
from the resource estimator.}
\label{fig:mem_decoded}
\end{figure}

\textbf{Decoded memory experiment.}
Beyond the round-trip circuits, we run a single-logical-qubit memory kernel
(prepare $|0_L\rangle$, syndrome rounds, measure) as a true QEC memory
experiment: the compiler emits the
Steane \steane{} Stim circuit (three syndrome rounds under depolarizing noise)
via \texttt{--ftqc-lower-to-stim}, decoded with PyMatching MWPM ($60{,}000$
shots/point).  Figure~\ref{fig:mem_decoded} plots the decoded $p_L$, the
undecoded rate, and the heuristic $p_{\rm total}$. MWPM roughly halves the error
at every $p$. The heuristic is optimistic below ${\sim}5{\times}10^{-3}$ (it
omits encoding and multi-round noise) and pessimistic above it (no active
decoding); this crossover is descriptive, not a calibration of the estimator.

\subsection{Hardware Targeting: IQM Pathfinder}
\label{sec:hardware}
We target ORNL's \emph{Pathfinder}, a 20-qubit IQM Radiance superconducting machine deployed in 2026~\cite{ornl2026pathfinder}.  The IQM back end (\texttt{--ftqc-lower-to-iqm-json}, or two-level \texttt{--ftqc-qphys-to-iqm-json}) emits IQM Resonance JSON in the native gate set (phased-$X$ rotations \texttt{prx}, \texttt{cz}, \texttt{measure}), so a Steane-encoded logical circuit becomes submission-ready.  Table~\ref{tab:iqm} reports the native-gate cost and physical-qubit footprint for one and two logical qubits.

\begin{table}[h]
\centering\small
\caption{\texttt{FTQC}$\rightarrow$IQM native-gate lowering of Steane-encoded circuits for the 20-qubit Pathfinder.  $n_L$: logical qubits; $n_\text{data}$: data qubits (the shared-ancilla two-level path adds three ancilla per logical qubit); \texttt{prx}/\texttt{cz}/\texttt{meas}: native-gate counts.}
\label{tab:iqm}
\vspace{-4pt}
\begin{tabular}{@{}lcccccc@{}}
\toprule
\textbf{Circuit} & $n_L$ & $n_\text{data}$ & \texttt{prx} & \texttt{cz} & \texttt{meas} & fit/20 \\
\midrule
logical $|0\rangle$, measure        & 1 & 7  & 42 & 9 & 7  & yes \\
logical $|0\rangle$, $H$, measure   & 1 & 7  & 56 & 9 & 7  & yes \\
Bell ($H\cdot$CNOT)                 & 2 & 14 & 126 & 25 & 14 & yes \\
\bottomrule
\end{tabular}
\end{table}

\noindent\textbf{Resource fit.} The encoded lowering (Table~\ref{tab:iqm}) places one logical Steane qubit on 7 data qubits, or 10 with its shared-ancilla block, out of Pathfinder's 20 qubits. 
Therefore, Pathfinder can run the full Steane encoder, transversal gates, and repeated syndrome extraction for quantum circuits consisting of two logical qubits such as a Bell circuit.
\textbf{Topology-aware layout.}  From Pathfinder's coupling map (20 qubits, 30 edges) and per-qubit \texttt{prx} calibration, a layout search picks the candidate minimizing total single-qubit error ($\sum\texttt{prx\_error}=1.3\times10^{-3}$), which \texttt{--ftqc-qphys-init-layout} stamps onto each \texttt{qphys.alloc}. The seven-qubit encoder's adjacency is not a device subgraph, so physical circuits are routed (SWAP insertion via \texttt{qiskit-iqm}'s \texttt{transpile\_to\_IQM}) before submission.
\textbf{Execution path.}  We use the Q-IRIS framework~\cite{hpcasia26qiris} as a backend runtime to streamline the overall process from FTQC compilation to execution on the IQM machine, where  \texttt{@iris.kernel} tasks drive QASM $\rightarrow$ logical MLIR $\rightarrow$ physical MLIR $\rightarrow$ IQM JSON $\rightarrow$ route $\rightarrow$ submit, after which syndrome decoding recovers the logical outcome. The emitted JSON is device-valid (native \texttt{prx}/\texttt{cz}/\texttt{measure} on \texttt{QB} loci) and runs on Pathfinder.  \textbf{On-device results.}  For one logical qubit (identity kernel, ideal logical $|0\rangle$, 512 shots), decoding the seven-qubit data readout lifts the logical-$|0\rangle$ fraction from $79.3\%$ (raw transversal parity) to $93.8\%$, a $3.3\times$ error reduction. For the two-logical Bell circuit (14 physical qubits, 512 shots), decoding raises $P(|00\rangle)+P(|11\rangle)$ from $57.6\%$ to $76.7\%$ and cuts $P(|01\rangle)+P(|10\rangle)$ from $42.3\%$ to $23.2\%$. This shows a proof-of-concept that the pipeline reaches hardware and that classical decoding of the encoded readout improves the logical statistics. They are single 512-shot runs without confidence intervals or an unencoded baseline, so they do \emph{not} establish fault-tolerant error suppression; a threshold demonstration would need code-distance and shot scaling.

\subsection{Circuit Depth and Critical-Path Analysis}
The SE tally is not circuit depth: it counts individual X- and Z-type extractions and ignores overlap between logical blocks. Grouping each X+Z pair into one cycle gives 2 cycles for Bell, 4 for GHZ-4, 10 for Grover, and 18 for QFT-3 before physical expansion. Actual depth also depends on stabilizer scheduling, distance, routing, and ancilla. QFT-3 further serializes its six $T$ states through one factory: seven cycles per output at $d=3$~\cite{litinski2019}, so 42 cycles for six outputs, or 21 with two parallel factories. These are scheduling estimates, not hardware-latency predictions.

\subsection{Compilation Performance}
\begin{figure}[t]
\centering
\includegraphics[width=\columnwidth]{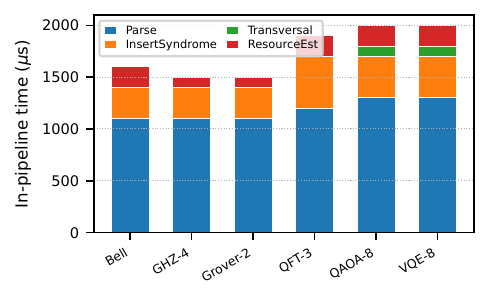}
\vspace{-1mm}
\caption{Per-pass in-pipeline compilation time (minimum of seven runs, \texttt{-mlir-timing}, AMD EPYC 7513 single core, \texttt{-{}-mlir-disable-threading}) for the six benchmarks (parametrized families at $n=8$). Parsing dominates; the FTQC passes together add under $1\,\text{ms}$. This in-pipeline cost excludes the one-time $\approx170\,\text{ms}$ process/library startup; times are quantized to $0.1\,\text{ms}$ by the timing report.}
\label{fig:passtime}
\end{figure}

Total wall-clock is dominated by a one-time $\approx170\,\text{ms}$ of library startup (dynamic loading and static initialization of the MLIR shared libraries), independent of circuit size; a leaner local build cuts this by an order of magnitude. We therefore report the \emph{in-pipeline} cost from \texttt{-mlir-timing} (minimum of seven runs), which excludes startup. Figure~\ref{fig:passtime} breaks it down for all six benchmarks (parametrized families at $n=8$): parsing dominates at $1.1$ to $1.3\,\text{ms}$ and grows with size, while the FTQC passes together add under $1\,\text{ms}$. Insert-syndrome is the largest ($0.2$ to $0.5\,\text{ms}$), transversal lowering is nearly free, and resource estimation is roughly constant ($\approx0.2\,\text{ms}$). Each pass is a single $O(n)$ walk over the IR, so in-pipeline compilation is parse-bound and grows linearly with gate count, quadratically for QFT (its controlled-phase ladder has $O(n^2)$ gates), as quantified in Section~\ref{sec:scaling}.

\section{Scaling Experiments}
\label{sec:scaling}


\def\coordsBellDataQubits{(2,14) (4,28) (6,42) (8,56) (10,70) (12,84) (14,98) (16,112) (18,126) (20,140) (22,154) (24,168) (26,182) (28,196) (30,210) (32,224) (34,238) (36,252) (38,266) (40,280) (42,294) (44,308) (46,322) (48,336) (50,350) (52,364) (54,378) (56,392) (58,406) (60,420) (62,434) (64,448) (66,462) (68,476) (70,490) (72,504) (74,518) (76,532) (78,546) (80,560) (82,574) (84,588) (86,602) (88,616) (90,630) (92,644) (94,658) (96,672) (98,686) (100,700)}
\def\coordsBellAncillaQubits{(2,12) (4,24) (6,36) (8,48) (10,60) (12,72) (14,84) (16,96) (18,108) (20,120) (22,132) (24,144) (26,156) (28,168) (30,180) (32,192) (34,204) (36,216) (38,228) (40,240) (42,252) (44,264) (46,276) (48,288) (50,300) (52,312) (54,324) (56,336) (58,348) (60,360) (62,372) (64,384) (66,396) (68,408) (70,420) (72,432) (74,444) (76,456) (78,468) (80,480) (82,492) (84,504) (86,516) (88,528) (90,540) (92,552) (94,564) (96,576) (98,588) (100,600)}
\def\coordsBellDistilQubits{(2,0) (4,0) (6,0) (8,0) (10,0) (12,0) (14,0) (16,0) (18,0) (20,0) (22,0) (24,0) (26,0) (28,0) (30,0) (32,0) (34,0) (36,0) (38,0) (40,0) (42,0) (44,0) (46,0) (48,0) (50,0) (52,0) (54,0) (56,0) (58,0) (60,0) (62,0) (64,0) (66,0) (68,0) (70,0) (72,0) (74,0) (76,0) (78,0) (80,0) (82,0) (84,0) (86,0) (88,0) (90,0) (92,0) (94,0) (96,0) (98,0) (100,0)}
\def\coordsBellPhysTotal{(2,26) (4,52) (6,78) (8,104) (10,130) (12,156) (14,182) (16,208) (18,234) (20,260) (22,286) (24,312) (26,338) (28,364) (30,390) (32,416) (34,442) (36,468) (38,494) (40,520) (42,546) (44,572) (46,598) (48,624) (50,650) (52,676) (54,702) (56,728) (58,754) (60,780) (62,806) (64,832) (66,858) (68,884) (70,910) (72,936) (74,962) (76,988) (78,1014) (80,1040) (82,1066) (84,1092) (86,1118) (88,1144) (90,1170) (92,1196) (94,1222) (96,1248) (98,1274) (100,1300)}
\def\coordsBellECRounds{(2,4) (4,8) (6,12) (8,16) (10,20) (12,24) (14,28) (16,32) (18,36) (20,40) (22,44) (24,48) (26,52) (28,56) (30,60) (32,64) (34,68) (36,72) (38,76) (40,80) (42,84) (44,88) (46,92) (48,96) (50,100) (52,104) (54,108) (56,112) (58,116) (60,120) (62,124) (64,128) (66,132) (68,136) (70,140) (72,144) (74,148) (76,152) (78,156) (80,160) (82,164) (84,168) (86,172) (88,176) (90,180) (92,184) (94,188) (96,192) (98,196) (100,200)}
\def\coordsBellTGates{(2,0) (4,0) (6,0) (8,0) (10,0) (12,0) (14,0) (16,0) (18,0) (20,0) (22,0) (24,0) (26,0) (28,0) (30,0) (32,0) (34,0) (36,0) (38,0) (40,0) (42,0) (44,0) (46,0) (48,0) (50,0) (52,0) (54,0) (56,0) (58,0) (60,0) (62,0) (64,0) (66,0) (68,0) (70,0) (72,0) (74,0) (76,0) (78,0) (80,0) (82,0) (84,0) (86,0) (88,0) (90,0) (92,0) (94,0) (96,0) (98,0) (100,0)}
\def\coordsBellTimeUs{(2,1800.0) (4,1900.0) (6,2000.0) (8,2200.0) (10,2200.0) (12,2100.0) (14,2200.0) (16,2500.0) (18,2200.0) (20,2300.0) (22,2500.0) (24,2300.0) (26,2600.0) (28,2600.0) (30,2600.0) (32,2700.0) (34,2800.0) (36,2900.0) (38,2900.0) (40,2800.0) (42,3000.0) (44,2900.0) (46,2900.0) (48,2900.0) (50,3300.0) (52,3100.0) (54,3700.0) (56,3400.0) (58,3900.0) (60,3500.0) (62,3400.0) (64,3700.0) (66,3600.0) (68,3600.0) (70,4000.0) (72,3900.0) (74,4000.0) (76,3800.0) (78,4300.0) (80,4100.0) (82,4200.0) (84,3900.0) (86,5700.0) (88,4200.0) (90,4000.0) (92,4200.0) (94,4100.0) (96,4200.0) (98,4400.0) (100,4200.0)}

\def\coordsGHZDataQubits{(2,14) (4,28) (6,42) (8,56) (10,70) (12,84) (14,98) (16,112) (18,126) (20,140) (22,154) (24,168) (26,182) (28,196) (30,210) (32,224) (34,238) (36,252) (38,266) (40,280) (42,294) (44,308) (46,322) (48,336) (50,350) (52,364) (54,378) (56,392) (58,406) (60,420) (62,434) (64,448) (66,462) (68,476) (70,490) (72,504) (74,518) (76,532) (78,546) (80,560) (82,574) (84,588) (86,602) (88,616) (90,630) (92,644) (94,658) (96,672) (98,686) (100,700)}
\def\coordsGHZAncillaQubits{(2,12) (4,24) (6,36) (8,48) (10,60) (12,72) (14,84) (16,96) (18,108) (20,120) (22,132) (24,144) (26,156) (28,168) (30,180) (32,192) (34,204) (36,216) (38,228) (40,240) (42,252) (44,264) (46,276) (48,288) (50,300) (52,312) (54,324) (56,336) (58,348) (60,360) (62,372) (64,384) (66,396) (68,408) (70,420) (72,432) (74,444) (76,456) (78,468) (80,480) (82,492) (84,504) (86,516) (88,528) (90,540) (92,552) (94,564) (96,576) (98,588) (100,600)}
\def\coordsGHZDistilQubits{(2,0) (4,0) (6,0) (8,0) (10,0) (12,0) (14,0) (16,0) (18,0) (20,0) (22,0) (24,0) (26,0) (28,0) (30,0) (32,0) (34,0) (36,0) (38,0) (40,0) (42,0) (44,0) (46,0) (48,0) (50,0) (52,0) (54,0) (56,0) (58,0) (60,0) (62,0) (64,0) (66,0) (68,0) (70,0) (72,0) (74,0) (76,0) (78,0) (80,0) (82,0) (84,0) (86,0) (88,0) (90,0) (92,0) (94,0) (96,0) (98,0) (100,0)}
\def\coordsGHZPhysTotal{(2,26) (4,52) (6,78) (8,104) (10,130) (12,156) (14,182) (16,208) (18,234) (20,260) (22,286) (24,312) (26,338) (28,364) (30,390) (32,416) (34,442) (36,468) (38,494) (40,520) (42,546) (44,572) (46,598) (48,624) (50,650) (52,676) (54,702) (56,728) (58,754) (60,780) (62,806) (64,832) (66,858) (68,884) (70,910) (72,936) (74,962) (76,988) (78,1014) (80,1040) (82,1066) (84,1092) (86,1118) (88,1144) (90,1170) (92,1196) (94,1222) (96,1248) (98,1274) (100,1300)}
\def\coordsGHZECRounds{(2,4) (4,8) (6,12) (8,16) (10,20) (12,24) (14,28) (16,32) (18,36) (20,40) (22,44) (24,48) (26,52) (28,56) (30,60) (32,64) (34,68) (36,72) (38,76) (40,80) (42,84) (44,88) (46,92) (48,96) (50,100) (52,104) (54,108) (56,112) (58,116) (60,120) (62,124) (64,128) (66,132) (68,136) (70,140) (72,144) (74,148) (76,152) (78,156) (80,160) (82,164) (84,168) (86,172) (88,176) (90,180) (92,184) (94,188) (96,192) (98,196) (100,200)}
\def\coordsGHZTGates{(2,0) (4,0) (6,0) (8,0) (10,0) (12,0) (14,0) (16,0) (18,0) (20,0) (22,0) (24,0) (26,0) (28,0) (30,0) (32,0) (34,0) (36,0) (38,0) (40,0) (42,0) (44,0) (46,0) (48,0) (50,0) (52,0) (54,0) (56,0) (58,0) (60,0) (62,0) (64,0) (66,0) (68,0) (70,0) (72,0) (74,0) (76,0) (78,0) (80,0) (82,0) (84,0) (86,0) (88,0) (90,0) (92,0) (94,0) (96,0) (98,0) (100,0)}
\def\coordsGHZTimeUs{(2,1900.0) (4,2000.0) (6,2100.0) (8,2100.0) (10,2200.0) (12,2100.0) (14,2200.0) (16,2300.0) (18,2300.0) (20,2500.0) (22,2400.0) (24,2400.0) (26,2400.0) (28,2600.0) (30,2800.0) (32,2700.0) (34,2600.0) (36,2800.0) (38,3000.0) (40,2900.0) (42,3100.0) (44,3600.0) (46,3100.0) (48,3100.0) (50,3200.0) (52,3200.0) (54,3300.0) (56,3500.0) (58,3800.0) (60,3800.0) (62,3800.0) (64,3700.0) (66,3800.0) (68,3900.0) (70,3900.0) (72,3700.0) (74,4000.0) (76,4000.0) (78,4000.0) (80,4200.0) (82,4100.0) (84,4200.0) (86,4800.0) (88,4300.0) (90,4400.0) (92,4400.0) (94,4400.0) (96,4500.0) (98,4400.0) (100,4800.0)}

\def\coordsGroverDataQubits{(2,50) (4,100) (6,150) (8,200) (10,250) (12,300) (14,350) (16,400) (18,450) (20,500) (22,550) (24,600) (26,650) (28,700) (30,750) (32,800) (34,850) (36,900) (38,950) (40,1000) (42,1050) (44,1100) (46,1150) (48,1200) (50,1250) (52,1300) (54,1350) (56,1400) (58,1450) (60,1500) (62,1550) (64,1600) (66,1650) (68,1700) (70,1750) (72,1800) (74,1850) (76,1900) (78,1950) (80,2000) (82,2050) (84,2100) (86,2150) (88,2200) (90,2250) (92,2300) (94,2350) (96,2400) (98,2450) (100,2500)}
\def\coordsGroverAncillaQubits{(2,48) (4,96) (6,144) (8,192) (10,240) (12,288) (14,336) (16,384) (18,432) (20,480) (22,528) (24,576) (26,624) (28,672) (30,720) (32,768) (34,816) (36,864) (38,912) (40,960) (42,1008) (44,1056) (46,1104) (48,1152) (50,1200) (52,1248) (54,1296) (56,1344) (58,1392) (60,1440) (62,1488) (64,1536) (66,1584) (68,1632) (70,1680) (72,1728) (74,1776) (76,1824) (78,1872) (80,1920) (82,1968) (84,2016) (86,2064) (88,2112) (90,2160) (92,2208) (94,2256) (96,2304) (98,2352) (100,2400)}
\def\coordsGroverDistilQubits{(2,0) (4,0) (6,0) (8,0) (10,0) (12,0) (14,0) (16,0) (18,0) (20,0) (22,0) (24,0) (26,0) (28,0) (30,0) (32,0) (34,0) (36,0) (38,0) (40,0) (42,0) (44,0) (46,0) (48,0) (50,0) (52,0) (54,0) (56,0) (58,0) (60,0) (62,0) (64,0) (66,0) (68,0) (70,0) (72,0) (74,0) (76,0) (78,0) (80,0) (82,0) (84,0) (86,0) (88,0) (90,0) (92,0) (94,0) (96,0) (98,0) (100,0)}
\def\coordsGroverPhysTotal{(2,98) (4,196) (6,294) (8,392) (10,490) (12,588) (14,686) (16,784) (18,882) (20,980) (22,1078) (24,1176) (26,1274) (28,1372) (30,1470) (32,1568) (34,1666) (36,1764) (38,1862) (40,1960) (42,2058) (44,2156) (46,2254) (48,2352) (50,2450) (52,2548) (54,2646) (56,2744) (58,2842) (60,2940) (62,3038) (64,3136) (66,3234) (68,3332) (70,3430) (72,3528) (74,3626) (76,3724) (78,3822) (80,3920) (82,4018) (84,4116) (86,4214) (88,4312) (90,4410) (92,4508) (94,4606) (96,4704) (98,4802) (100,4900)}
\def\coordsGroverECRounds{(2,20) (4,40) (6,60) (8,80) (10,100) (12,120) (14,140) (16,160) (18,180) (20,200) (22,220) (24,240) (26,260) (28,280) (30,300) (32,320) (34,340) (36,360) (38,380) (40,400) (42,420) (44,440) (46,460) (48,480) (50,500) (52,520) (54,540) (56,560) (58,580) (60,600) (62,620) (64,640) (66,660) (68,680) (70,700) (72,720) (74,740) (76,760) (78,780) (80,800) (82,820) (84,840) (86,860) (88,880) (90,900) (92,920) (94,940) (96,960) (98,980) (100,1000)}
\def\coordsGroverTGates{(2,0) (4,0) (6,0) (8,0) (10,0) (12,0) (14,0) (16,0) (18,0) (20,0) (22,0) (24,0) (26,0) (28,0) (30,0) (32,0) (34,0) (36,0) (38,0) (40,0) (42,0) (44,0) (46,0) (48,0) (50,0) (52,0) (54,0) (56,0) (58,0) (60,0) (62,0) (64,0) (66,0) (68,0) (70,0) (72,0) (74,0) (76,0) (78,0) (80,0) (82,0) (84,0) (86,0) (88,0) (90,0) (92,0) (94,0) (96,0) (98,0) (100,0)}
\def\coordsGroverTimeUs{(2,2000.0) (4,2300.0) (6,2400.0) (8,2600.0) (10,2800.0) (12,2900.0) (14,3200.0) (16,3300.0) (18,3500.0) (20,3900.0) (22,4300.0) (24,4600.0) (26,4800.0) (28,4900.0) (30,5100.0) (32,5300.0) (34,5500.0) (36,5100.0) (38,5900.0) (40,5800.0) (42,5700.0) (44,5800.0) (46,9500.0) (48,7300.0) (50,7300.0) (52,6500.0) (54,7100.0) (56,7500.0) (58,7200.0) (60,7400.0) (62,11200.0) (64,7600.0) (66,8500.0) (68,8100.0) (70,8600.0) (72,8500.0) (74,8600.0) (76,8700.0) (78,9000.0) (80,9000.0) (82,9300.0) (84,9400.0) (86,9700.0) (88,9900.0) (90,10000.0) (92,10300.0) (94,10500.0) (96,10900.0) (98,11100.0) (100,11100.0)}

\def\coordsQFTDataQubits{(2,14) (3,21) (4,28) (5,35) (6,42) (7,49) (8,56) (9,63) (10,70) (11,77) (12,84) (13,91) (14,98) (15,105) (16,112) (17,119) (18,126) (19,133) (20,140) (21,147) (22,154) (23,161) (24,168) (25,175) (26,182) (27,189) (28,196) (29,203) (30,210) (31,217) (32,224) (33,231) (34,238) (35,245) (36,252) (37,259) (38,266) (39,273) (40,280) (41,287) (42,294) (43,301) (44,308) (45,315) (46,322) (47,329) (48,336) (49,343) (50,350) (51,357)}
\def\coordsQFTAncillaQubits{(2,12) (3,18) (4,24) (5,30) (6,36) (7,42) (8,48) (9,54) (10,60) (11,66) (12,72) (13,78) (14,84) (15,90) (16,96) (17,102) (18,108) (19,114) (20,120) (21,126) (22,132) (23,138) (24,144) (25,150) (26,156) (27,162) (28,168) (29,174) (30,180) (31,186) (32,192) (33,198) (34,204) (35,210) (36,216) (37,222) (38,228) (39,234) (40,240) (41,246) (42,252) (43,258) (44,264) (45,270) (46,276) (47,282) (48,288) (49,294) (50,300) (51,306)}
\def\coordsQFTDistilQubits{(2,195) (3,195) (4,195) (5,195) (6,195) (7,195) (8,195) (9,195) (10,195) (11,195) (12,195) (13,195) (14,195) (15,195) (16,195) (17,195) (18,195) (19,195) (20,195) (21,195) (22,195) (23,195) (24,195) (25,195) (26,195) (27,195) (28,195) (29,195) (30,195) (31,195) (32,195) (33,195) (34,195) (35,195) (36,195) (37,195) (38,195) (39,195) (40,195) (41,195) (42,195) (43,195) (44,195) (45,195) (46,195) (47,195) (48,195) (49,195) (50,195) (51,195)}
\def\coordsQFTPhysTotal{(2,221) (3,234) (4,247) (5,260) (6,273) (7,286) (8,299) (9,312) (10,325) (11,338) (12,351) (13,364) (14,377) (15,390) (16,403) (17,416) (18,429) (19,442) (20,455) (21,468) (22,481) (23,494) (24,507) (25,520) (26,533) (27,546) (28,559) (29,572) (30,585) (31,598) (32,611) (33,624) (34,637) (35,650) (36,663) (37,676) (38,689) (39,702) (40,715) (41,728) (42,741) (43,754) (44,767) (45,780) (46,793) (47,806) (48,819) (49,832) (50,845) (51,858)}
\def\coordsQFTECRounds{(2,18) (3,36) (4,68) (5,102) (6,150) (7,200) (8,264) (9,330) (10,410) (11,492) (12,588) (13,686) (14,798) (15,912) (16,1040) (17,1170) (18,1314) (19,1460) (20,1620) (21,1782) (22,1958) (23,2136) (24,2328) (25,2522) (26,2730) (27,2940) (28,3164) (29,3390) (30,3630) (31,3872) (32,4128) (33,4386) (34,4658) (35,4932) (36,5220) (37,5510) (38,5814) (39,6120) (40,6440) (41,6762) (42,7098) (43,7436) (44,7788) (45,8142) (46,8510) (47,8880) (48,9264) (49,9650) (50,10050) (51,10452)}
\def\coordsQFTTGates{(2,2) (3,6) (4,12) (5,20) (6,30) (7,42) (8,56) (9,72) (10,90) (11,110) (12,132) (13,156) (14,182) (15,210) (16,240) (17,272) (18,306) (19,342) (20,380) (21,420) (22,462) (23,506) (24,552) (25,600) (26,650) (27,702) (28,756) (29,812) (30,870) (31,930) (32,992) (33,1056) (34,1122) (35,1190) (36,1260) (37,1332) (38,1406) (39,1482) (40,1560) (41,1640) (42,1722) (43,1806) (44,1892) (45,1980) (46,2070) (47,2162) (48,2256) (49,2352) (50,2450) (51,2550)}
\def\coordsQFTTimeUs{(2,2000.0) (3,2100.0) (4,2500.0) (5,2700.0) (6,3400.0) (7,3600.0) (8,4300.0) (9,4800.0) (10,5400.0) (11,6100.0) (12,7000.0) (13,8200.0) (14,9400.0) (15,10200.0) (16,11400.0) (17,13300.0) (18,15000.0) (19,14900.0) (20,16400.0) (21,17800.0) (22,19300.0) (23,21200.0) (24,22900.0) (25,25200.0) (26,28400.0) (27,28800.0) (28,31500.0) (29,33400.0) (30,34800.0) (31,37100.0) (32,40200.0) (33,42100.0) (34,44700.0) (35,48400.0) (36,50300.0) (37,53000.0) (38,55300.0) (39,57800.0) (40,60600.0) (41,64199.99999999999) (42,69400.0) (43,72700.0) (44,73500.0) (45,76800.0) (46,80100.0) (47,84600.0) (48,89000.0) (49,91200.0) (50,96700.0) (51,100900.0)}

\def\coordsQAOADataQubits{(2,14) (4,28) (6,42) (8,56) (10,70) (12,84) (14,98) (16,112) (18,126) (20,140) (22,154) (24,168) (26,182) (28,196) (30,210) (32,224)}
\def\coordsQAOAAncillaQubits{(2,12) (4,24) (6,36) (8,48) (10,60) (12,72) (14,84) (16,96) (18,108) (20,120) (22,132) (24,144) (26,156) (28,168) (30,180) (32,192)}
\def\coordsQAOADistilQubits{(2,195) (4,195) (6,195) (8,195) (10,195) (12,195) (14,195) (16,195) (18,195) (20,195) (22,195) (24,195) (26,195) (28,195) (30,195) (32,195)}
\def\coordsQAOAPhysTotal{(2,221) (4,247) (6,273) (8,299) (10,325) (12,351) (14,377) (16,403) (18,429) (20,455) (22,481) (24,507) (26,533) (28,559) (30,585) (32,611)}
\def\coordsQAOAECRounds{(2,16) (4,32) (6,48) (8,64) (10,80) (12,96) (14,112) (16,128) (18,144) (20,160) (22,176) (24,192) (26,208) (28,224) (30,240) (32,256)}
\def\coordsQAOATGates{(2,4) (4,8) (6,12) (8,16) (10,20) (12,24) (14,28) (16,32) (18,36) (20,40) (22,44) (24,48) (26,52) (28,56) (30,60) (32,64)}
\def\coordsQAOATimeUs{(2,2000.0) (4,2700.0) (6,2700.0) (8,2500.0) (10,3000.0) (12,3700.0) (14,4000.0) (16,4200.0) (18,4400.0) (20,4700.0) (22,5100.0) (24,5800.0) (26,4100.0) (28,4000.0) (30,4500.0) (32,4300.0)}

\def\coordsVQEDataQubits{(2,14) (4,28) (6,42) (8,56) (10,70) (12,84) (14,98) (16,112) (18,126) (20,140) (22,154) (24,168) (26,182) (28,196) (30,210) (32,224)}
\def\coordsVQEAncillaQubits{(2,12) (4,24) (6,36) (8,48) (10,60) (12,72) (14,84) (16,96) (18,108) (20,120) (22,132) (24,144) (26,156) (28,168) (30,180) (32,192)}
\def\coordsVQEDistilQubits{(2,195) (4,195) (6,195) (8,195) (10,195) (12,195) (14,195) (16,195) (18,195) (20,195) (22,195) (24,195) (26,195) (28,195) (30,195) (32,195)}
\def\coordsVQEPhysTotal{(2,221) (4,247) (6,273) (8,299) (10,325) (12,351) (14,377) (16,403) (18,429) (20,455) (22,481) (24,507) (26,533) (28,559) (30,585) (32,611)}
\def\coordsVQEECRounds{(2,10) (4,22) (6,34) (8,46) (10,58) (12,70) (14,82) (16,94) (18,106) (20,118) (22,130) (24,142) (26,154) (28,166) (30,178) (32,190)}
\def\coordsVQETGates{(2,4) (4,8) (6,12) (8,16) (10,20) (12,24) (14,28) (16,32) (18,36) (20,40) (22,44) (24,48) (26,52) (28,56) (30,60) (32,64)}
\def\coordsVQETimeUs{(2,2200.0) (4,2600.0) (6,2700.0) (8,2400.0) (10,2400.0) (12,2800.0) (14,3000.0) (16,3500.0) (18,4100.0) (20,3200.0) (22,3300.0) (24,4900.0) (26,3600.0) (28,3700.0) (30,4400.0) (32,4100.0)}

We generate six circuit families over $n$ logical qubits (Bell,
GHZ, Grover to $n=100$; QFT to $n=51$, limited by its quadratic gate count;
QAOA and VQE to $n=32$). We compile each with
\texttt{-{}-ftqc-insert-syndrome} \texttt{-{}-ftqc-transversal-gates}
\texttt{-{}-ftqc-resource-estimate} and record the \emph{in-pipeline} compile time
from \texttt{-mlir-timing} (minimum of three runs on an AMD EPYC 7513 pinned to a
single core, MLIR threading disabled, $-O2$), which excludes one-time process
and library startup. We also parse the physical-qubit breakdown from the
resource annotation. Bell, GHZ, QFT, QAOA, and VQE use Steane $d=3$; Grover-2
uses surface $d=5$. These generated families are compiler-throughput
microbenchmarks, not evidence from application-scale programs. The Bell family
repeats a small Clifford pattern to measure linear IR growth. We do not treat
it as a distinct large algorithm.

\begin{figure}[t]
\centering
\includegraphics[width=1.0\columnwidth]{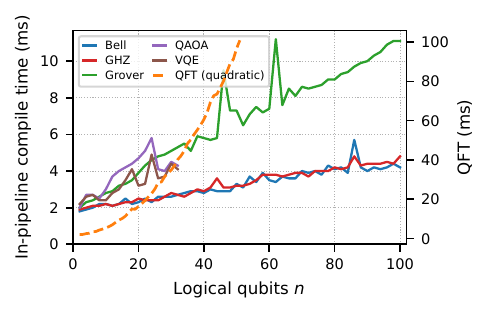}
\vspace{-1mm}
\caption{In-pipeline compilation time (parse + passes, \texttt{-mlir-timing}, minimum of three runs, AMD EPYC 7513 single core, threading disabled), excluding the one-time $\approx170\,\text{ms}$ process/library startup. The near-linear families (Bell, GHZ, QAOA, VQE, left axis) stay under $6\,\text{ms}$; Grover reaches $\approx11\,\text{ms}$ at $n = 100$; QFT (right axis) grows quadratically to $\approx100\,\text{ms}$ at $n = 51$, tracking its $O(n^2)$ controlled-phase ladder.}
\label{fig:compile_time}
\vspace{-0.15in}
\end{figure}

\noindent\textbf{Findings.}
Total wall-clock is dominated by a constant $\approx170$\,ms of process/library
startup on this node. The \emph{in-pipeline} time (parse + passes, excluding
startup) is parse-bound: each added gate costs well under one millisecond, so it
grows linearly in gate count. This gives $O(n)$ for the near-Clifford families
(under $6$\,ms at $n=100$ for Bell and GHZ) and $O(n^2)$ for QFT (about
$100$\,ms at $n=51$), tracking its quadratic gate count
(Figure~\ref{fig:compile_time}). The physical-qubit demand grows as $13n$
(Steane $d=3$) and $49n$ (surface $d=5$) for the purely Clifford workloads. QFT
adds a fixed 195-qubit distillation penalty for any $n\ge2$ (one 15-to-1 factory
serializing all $T$ gates), and both its $T$-count and syndrome-extraction tally scale as
$\Theta(n^2)$. The dominant levers for large-scale fault-tolerant QFT are thus
$T$-gate reduction and depth-optimal factory scheduling.

\section{Limitations and Scope}
\label{sec:limitations}
Our evaluation focuses on the compiler layer, while several extensions remain for future work. The reported logical-error values use a leading-order model for approximate resource
  planning. They should not be interpreted as bounds calibrated to a specific decoder. Likewise, the distillation estimate measures peak qubit use rather than space-time cost. A
  complete factory model and scheduling method are left for future work. Stim supports stabilizer circuits, so we replace non-Clifford gates with Clifford stand-ins during
  simulation. These experiments evaluate noisy syndrome extraction and decoding, but they do not demonstrate end-to-end fault tolerance. The benchmarks cover several circuit
  structures, although they remain modest in size. Future studies should include larger standard workloads and direct comparisons with external resource estimators. The IR could also
  be extended to support dynamic control, routing, and scheduling. Our hardware experiments provide initial evidence that the compiler can target a physical device and that decoding
  improves the measured output statistics. Demonstrating threshold behavior would require experiments across multiple code distances and substantially more shots. These limitations
  do not change the paper’s core contribution: a typed, ECC-aware IR with resource analysis integrated into the compilation pipeline. Instead, they identify clear directions for
  further development.


\section{Related Work}
\label{sec:related}

\textbf{Quantum compiler frameworks.} Qiskit~\cite{qiskit} and Cirq~\cite{cirq} provide extensive gate libraries and NISQ optimization passes, including transpilation to native instruction sets and noise-aware routing. They offer no dedicated representation for ECC types, syndrome-extraction operations, or magic-state distillation factories. tket~\cite{tket} targets retargetable compilation with Clifford optimization and routing, but models no logical-level fault-tolerant abstractions. Earlier LLVM-based infrastructures such as ScaffCC~\cite{scaffcc2015} and XACC~\cite{xacc2019} built scalable logical-level workflows without typed IR constructs for QEC metadata or fault-tolerant resources.

\textbf{MLIR-based and MLIR-inspired quantum frameworks.} The PennyLane/Catalyst quantum dialect~\cite{bergholm2018} and the CUDA-Q Quake dialect from NVIDIA~\cite{cudaq,quakeDialect} model quantum-classical programs and physical or circuit-level gates within MLIR. QIR serves lower-level hardware targeting~\cite{qir2021}. The MQT Compiler Collection is closer architecturally: its imperative QC and value-semantic QCO dialects support interchange, structured control flow, circuit optimization, hardware mapping, and QIR lowering~\cite{mqtcompiler}. Like \texttt{FTQC}, it adopts MLIR progressive lowering, but the presented dialects do not model code-parameterized logical qubits or first-class fault-tolerant resources. Kirin instead provides Python-native, MLIR-inspired infrastructure for scientific embedded domain-specific languages; its documented QASM~2 and Stim dialects support analysis, rewriting, interpretation, and circuit-level QEC workflows~\cite{kirinrepo}. However, \texttt{FTQC} \textit{specializes} the logical fault-tolerant layer, with its LLVM/MLIR types carrying ECC family and distance, and its passes lower syndrome, magic-state, and Pauli-frame operations to physical circuits, ECC-derived resource estimates, simulators, interchange formats, and hardware targets.

\textbf{Benchmarking and ecosystem tooling.} The Munich Quantum Toolkit contributes MQT Bench~\cite{mqtbench2023}, a cross-level benchmark suite that improves comparability and reproducibility across quantum software tools. FTCircuitBench~\cite{harkness2026ftcircuitbench} extends this to fault-tolerant compilation with benchmarks pre-compiled in Clifford+$T$ and Pauli-Based Computation models. It supplies a shared workload baseline but, like other such suites, offers no typed fault-tolerant IR with operations for encoding, correction, distillation, and syndrome extraction.

\textbf{Fault-tolerant compilers.} Jabalizer~\cite{jabalizer2022} translates arbitrary circuits into Clifford+$T$ via graph-state synthesis, targeting surface-code lattice surgery. Stim~\cite{gidney2021stim} is a high-performance stabilizer-circuit simulator for syndrome decoding. Neither exposes an MLIR with composable passes. Lattice-surgery frameworks, including Litinski’s architecture-level formulation~\cite{litinski2019} and the high-throughput compiler of Watkins et al.~\cite{watkins2024}, give strong scheduling and resource analysis for large fault-tolerant workloads. NWQEC~\cite{nwqec2025, wangftcb2024, wangftcb2025} converts OpenQASM~2.0 to Clifford+$T$ via gridsynth and applies Pauli-Based Computation optimizations, but works at the gate-count layer rather than as a typed, ECC-aware IR. \texttt{FTQC} complements these tools with a type-safe, extensible MLIR that expresses multiple ECC families and custom transformation pipelines.

\textbf{Decoding and control-stack integration.} Riverlane’s work on parallel-window decoding~\cite{skoric2023} and real-time hardware-efficient decoding~\cite{barber2025} shows how decoder throughput and latency govern scalable fault-tolerant operation. These efforts are decoder- and control-stack-oriented, whereas \texttt{FTQC} targets compiler-level representation and transformation; together they point to a full-stack co-design path from typed fault-tolerant IR to real-time QEC.

\textbf{Resource estimators.} Microsoft’s Azure Quantum Resource Estimator~\cite{azureQRE} evaluates physical-qubit budgets for fault-tolerant algorithms but exposes no transformable IR. \texttt{FTQC} integrates resource estimation into the pipeline as a pass, enabling optimization on concrete resource metrics.

\texttt{FTQC} fills a gap in this landscape: a typed, SSA-based MLIR that exposes fault-tolerant abstractions (ECC types, syndrome-extraction operations, and magic-state values) as transformable IR within the standard MLIR pass infrastructure.


\section{Conclusion}
\label{sec:conclusion}

We presented \texttt{FTQC}, an open-source MLIR dialect for fault-tolerant quantum compilation. Its type-safe SSA representation covers logical qubits, syndromes, magic states, and Pauli frames. The dialect provides 31 operations and 18 passes. It emits OpenQASM~3, QIR, Stim, and native IQM instructions. Resource analysis reads Steane, surface-code, and color-code metadata directly from the IR.
The evaluation covers six circuit families. Logical simulations check algorithm behavior, while Steane-code experiments study noisy syndrome extraction with MWPM decoding. The latter use Clifford stand-ins for non-Clifford gates and do not establish fault tolerance. The largest noisy-QEC circuits contain 104 physical qubits. In-pipeline compilation of each reported benchmark takes under 2 ms, excluding startup. The Pathfinder experiments further show that the generated physical circuits can reach hardware and that classical decoding improves the observed output statistics.
Under the peak-space model, QFT-3's serially reused distillation factory requires 195 qubits, compared with 39 qubits for the encoded circuit. This result identifies factory scheduling as an important next step. A broader assessment also requires larger workloads, external baselines, adaptive QIR support, code-distance scaling, and a space-time resource model.

\pagebreak

\bibliographystyle{plain}
\bibliography{papers}

\end{document}